\documentclass[%
reprint,
 amsmath,amssymb,
 aps,
pra,
]{revtex4-1}

\usepackage{graphicx}
\usepackage{dcolumn}
\usepackage{bm}
\usepackage{layouts}
\usepackage{float}
\usepackage[dvipdfm,colorlinks,linkcolor=blue,anchorcolor=blue,citecolor=blue,urlcolor=blue]{hyperref}

\begin{document}

\preprint{APS/123-QED}

\title{Magic Wavelengths of the Yb \(6s^2\,{}^1S_0-6s6p\,{}^3P_1\) Intercombination Transition}

\author{T. A. Zheng$^{1}$}
\author{Y. A. Yang$^{1}$}%
\author{M. S. Safronova$^{2,3}$}
\author{U. I. Safronova$^{4}$}
\author{Zhuan-Xian Xiong$^{5}$}%
\author{T. Xia$^{1}$}%
 \email{txia1@ustc.edu.cn}
\author{Z.-T. Lu$^{1}$}%
  \email{ztlu@ustc.edu.cn}
\affiliation{$^1$Hefei National Laboratory for Physical Sciences at the Microscale, CAS Center for Excellence in Quantum Information and Quantum Physics, University of Science and Technology of China, Hefei, Anhui, China, 230026}%
\affiliation{$^2$Department of Physics and Astronomy, University of Delaware, Newark, Delaware 19716}
\affiliation{$^3$Joint Quantum Institute, National Institute of Standards and Technology and the University of Maryland, Gaithersburg, Maryland, 20899-8410}
\affiliation{$^4$Physics Department, University of Nevada, Reno, Nevada 89557}
\affiliation{$^5$Key Laboratory of Atomic Frequency Standards, Innovation Academy for Precision Measurement Science and Technology, Chinese Academy of Sciences, Wuhan 430071, P. R. China}

\date{\today}

\begin{abstract}
We calculate and measure the magic wavelengths for the $6s^2{}\,^1S_0-6s6p\,{}^3P_1$ intercombination transition of the neutral ytterbium atom. The calculation is performed with the \textit{ab initio} configuration interaction (CI) + all-order method. The measurement is done with laser spectroscopy on cold atoms in an optical dipole trap. The magic wavelengths are determined to be 1035.68(4) nm for the $\pi$ transition ($\Delta m = 0$) and 1036.12(3) nm for the $\sigma$ transitions ($|\Delta m| = 1$) in agreement with the calculated values. Laser cooling on the narrow intercombination transition could achieve better results for atoms in an optical dipole trap when the trap wavelength is tuned to near the magic wavelength.
\end{abstract}

\maketitle

\section{introduction} \label{sec:introduction}
For cold atoms in an optical dipole trap or an optical lattice, by tuning the wavelength of the trapping light to a “magic wavelength” \cite{ChaKohArn20,KatHasILI09,SasWilGri19,BarStaLem08,JiaJiaWu19}, the light shifts for the lower and upper states of a transition cancel each other so that the transition frequency is independent of the intensity of the trapping light. Indeed the transition frequency is unshifted as if the atoms were in free space. The optical lattice atomic clock based on this concept has achieved great metrological precision \cite{LudBoydYe15}. For quantum information and quantum communication applications \cite{LamNguLi18,JiaShuJun20,ZhaRobSaf11}, magic wavelength can significantly improve the quantum coherence time by reducing position-dependent and motional dephasing effects.

In this work, we focus on the magic wavelength for the $6s^2{}\,^1S_0-6s6p\,{}^3P_1$ intercombination transition of the neutral ytterbium atom (Yb). 

Yb is the choice for many cold-atom experiments including Bose-Einstein condensate (BEC) \cite{FukTakSug07,SugSeiYam11,YamMakKat13},  degenerate Fermi gas \cite{FukTakTak07,DorThoHun13}, optical lattice clock \cite{BelHinPhi14,HinShePhi13,LemLudBar09}, and quantum gas microscopy \cite{SasWilGri19,YamKobKun16,YamKobKat17,MirInoTam17,MirInoOku15}. For laser cooling of Yb, the narrow $6s^2{}\,^1S_0-6s6p\,{}^3P_1$ transition ($\Gamma/2\pi$ = 181 kHz) is widely used to achieve \(\mu\)K temperature. Knowing the magic wavelength and the gradient of differential ac Stark shifts near the magic wavelength, narrow-line cooling can be improved by employing Sisyphus cooling schemes \cite{SasWilGri19,CovJacMad19}.

In a previous work \cite{ZhiYanJun18}, the magic wavelength of $6s^2{}\,^1S_0-6s6p\,{}^3P_1$ was calculated using the configuration interaction (CI) + MBPT method. The results are 1035.7(2) nm for $\Delta m = 0$ and 1036.0(3) nm for $|\Delta m| = 1$. (Table~\ref{tab:table1})

\section{theory} \label{sec:theory}
Magic wavelengths for a specific atomic transition arise between the resonances of the relevant transitions originating from either the lower or upper state. The wavelength, or the corresponding frequency, can be calculated using the sum-over-states formula for a frequency-dependent polarizability of a state $v$ \cite{MitSafCla10}:
\begin{equation}
\alpha _{0}^{v}(\omega)=\frac{2}{3(2J+1)}\sum_{k}\frac{{\left\langle
k\left\| D\right\| v\right\rangle }^{2}(E_{k}-E_{v})}{(E_{k}-E_{v})^{2}-%
\omega ^{2}}.  \label{eq-pol}
\end{equation}
In the equation above, $J$ is the total angular moment of the state $v$,
${\left\langle k\left\|D\right\|v\right\rangle}$ is the reduced
electric-dipole matrix element, $\Delta E=E_k-E_v$
is the transition frequency, and $\omega$ is the light frequency, assumed to be at least several linewidths off the transition resonance. Light is assumed to have linear polarization in this calculation.

For the $6s^2{}\,^1S_0-6s6p\,{}^3P_1$ transition at 555.8 nm, there is a resonance for the polarizability of $6s6p\,{}^3P_1$ at 1032.45 nm \cite{NIST_ASD} due to the $6s6p\,{}^3P_1-5d6s\,{}^1D_2$ transition. The magic wavelength is determined as the crossing point of the polarizabilities of $6s^2{}\,^1S_0$ and $6s6p\,{}^3P_1$. The valence part of the polarizability is calculated by solving the inhomogeneous equation in the valence space
\begin{equation}
(E_v - H_{\textrm{eff}})|\Psi(v,m^{\prime})\rangle = D_{\mathrm{eff},q} |\Psi_0(v,J,m)\rangle
\label{eq1}
\end{equation}
for a state  $v$ with the total angular momentum $J$ and projection
$m$ \cite{PorRakKoz99a}. Here $H_{\textrm{eff}}$ includes the all-order corrections in the framework on the configuration interaction (CI) + all-order method \cite{SafPalJia09} which was used for the calculations of Yb blackbody shift and long-range interaction coefficients in \cite{SafPorCla12,PorSafDer14}. The effective dipole operator $D_{\textrm{eff}}$ includes random phase approximation (RPA) corrections.

In atomic units, polarizability has the dimension of volume, and its numerical values presented here are expressed in the unit of $a^3_0$, where $a_0\approx0.052918$ nm is the Bohr radius.

The total polarizability of the $^3P_1$ state is given by \cite{MitSafCla10}
\begin{equation}
\alpha=\alpha_0+\alpha_2 (\frac{3\cos^{2}\theta_{p}-1}{2})\frac{3m^2-J(J+1)}{J(2J-1)},
\end{equation}
where $\alpha_0$ and $\alpha_2$ are the scalar and tensor polarizability, respectively; $\theta_{p}$ is the angle between the quantization axis and the polarization of the light field. For all the following calculation, we assume $\theta_{p}=0$ for simplicity, thus the geometric factor $\frac{3\cos^{2}\theta_{p}-1}{2}=1$.
For $J=1$, we have  $\alpha=\alpha_0-2\alpha_2$ for $\Delta m=0$ and $\alpha=\alpha_0+\alpha_2$ for $|\Delta m|=1$.

\begin{figure}[t]
           \includegraphics[scale=0.32]{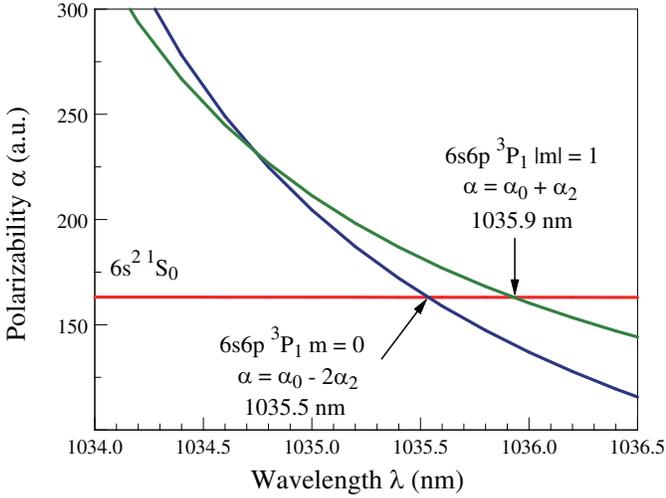}
            \caption{Calculated polarizabilities for $6s^2{}\,^1S_0$ and $6s6p\,^3P_1$, indicating the magic wavelengths for $6s^2{}\,^1S_0-6s6p\,{}^3P_1$ near the 1032.45 nm resonance.}
\label{MW_calculation}
\end{figure}

Our \textit{ab initio} CI + all-order calculations result in a resonance wavelength for the $6s6p\,^3P_1-5d6s\,^1D_2$ transition of 1026.36 nm, 0.6\% less than the experimental value of 1032.45 nm \cite{NIST_ASD}. We correct the placement of this resonance, using instead the measured value in the calculations. This is important since the magic wavelength is only 3 nm away from the resonance. We extract the contribution of the transition from $6s6p\,^3P_1$ (denoted $v$) to $5d6s\,^1D_2$ (denoted $k$) using the following formulae
\begin{eqnarray}
    \alpha_{0}&=&\frac{2}{3(2j_v+1)}\frac{{\left\langle k\left\|D\right\|v\right\rangle}^2(E_k-E_v)}{(E_k-E_v)^2-\omega^2}, \label{eq-1} \nonumber \\
    \alpha_{2}&=&-4C (-1)^{j_v+j_k+1}
            \left\{
                    \begin{array}{ccc}
                    j_v & 1 & j_k \\
                    1 & j_v & 2 \\
                    \end{array}
            \right\} \nonumber \\
      & &\times \frac{{\left\langle
            k\left\|D\right\|v\right\rangle}^2(E_k-E_v)}{
            (E_k-E_v)^2-\omega^2},
\end{eqnarray}
where $C$ is given by
\begin{equation}
            C =\left(\frac{5j_v(2j_v-1)}{6(j_v+1)(2j_v+1)(2j_v+3)}\right)^{1/2}. \nonumber
\end{equation}
The dynamic polarizabilities of Yb $6s^2{}\,^1S_0$ and $6s6p\,^3P_1$ states at a wavelength red-detuned from the 1032.45 nm resonance are plotted in Fig.~\ref{MW_calculation}. The resulting recommended values of the magic wavelengths are 1035.5 nm for $\Delta m$ = 0 and 1035.9 nm for $|\Delta m|$ = 1 (Table~\ref{tab:table1}).

\section {experimental setup} \label{sec:experimental setup}
\begin{figure}[b]
\includegraphics[width=3.4in]{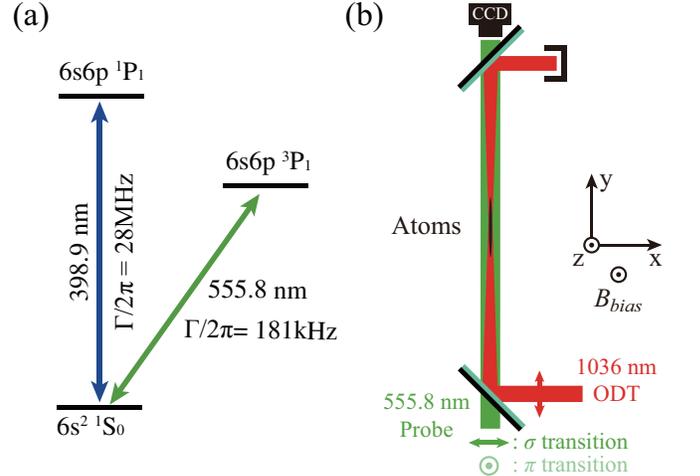}
\caption {\label{Apparatus} (a) Energy level structure showing the first-stage cooling transition at 398.9 nm and the second-stage cooling transition at 555.8 nm. (b) Setup for the magic wavelength measurement. The atoms in the optical dipole trap (ODT) of 1036 nm are under a uniform bias field $B_{bias}$ of 2 G. The 555.8 nm probe laser is for absorption spectroscopy, and its two perpendicular polarizations are for the detection of $\sigma$ and $\pi$ transition, respectively.}
\end{figure}

In the experiment (Fig.~\ref{Apparatus}), atoms are trapped in the first-stage blue magneto-optical trap (MOT) using the strong $6s^2\,{}^1S_0-6s6p\,{}^1P_1$ transition. The atoms are then transferred into the second-stage green MOT using the narrow-line $6s^2\,{}^1S_0-6s6p\,{}^3P_1$ transition. 

Doppler cooling on the narrow transition lowers the temperature of the atoms down to about 30
$\mu$K, as determined by time-of-flight measurements. The atoms are then transferred to an overlapping optical dipole trap (ODT) formed by a focused laser beam with a beam waist of 30 $\mu$m and a Raleigh length of 2.7 mm. At an ODT power of 30 W, the trap depth is 500 $\mu$K. The number of atoms in the ODT is about $1\times10^{6}$.

For transverse cooling, Zeeman slowing and the first-stage MOT, all on the $6s^2\,{}^1S_0-6s6p\,{}^1P_1$ transition, about 1 W of 398.9 nm light is produced by frequency doubling of 797.8 nm light from a diode-laser system. The 555.8 nm light for the second-stage MOT is produced in a similar way with an output power of about 200 mW. Both frequencies are stabilized to an ultra-low expansion cavity. Up to 50 W of laser power is available for the ODT generated by an Yb-doped fiber amplifier seeded with a diode laser.

\section{method} \label{sec:method}
We measure the differential ac Stark shift of $6s^2\,{}^1S_0-6s6p\,{}^3P_1$ transition and determine its magic wavelengths. The even isotope ${}^{174}$Yb is chosen for simplicity, since it has nuclear spin zero and thus no hyperfine structure. For experimental convenience, as shown in Fig.~\ref{Apparatus}(b), $B_{bias}$ is chosen to be in the z direction, and the ODT polarization in the x direction, so $\theta_{p}=\pi/2$. By measuring the shifts of both the $\pi$ ($\Delta m = 0$) and the $\sigma$ ($|\Delta m| = 1$) transitions, the scalar and tensor part of the ac Stark shift can be separated, and then recombined to yield shifts of $\theta_{p}=0$, or, for that matter, any $\theta_{p}$ values.

The differential ac Stark shift is proportional to the light intensity $I$ and the differential polarizability $\Delta \alpha$: $\triangle\nu\propto\triangle\alpha I$, where \(\triangle\alpha=\alpha^{e}-\alpha^{g}=\triangle\alpha_0-\frac{1}{2}(3m^2-2)\alpha_2^e\) (\(\theta_{p}=\frac{\pi}{2}\) in our experiment). We denote \({\rm DSSC}=\triangle\nu/P\) as the differential ac Stark shift coefficient (DSSC), where \(P\) is the ODT laser power. DSSC is measured at various trap laser wavelengths to determine the magic wavelength. At a magic wavelength, both DSSC and \(\triangle\alpha\) go to zero. 

\begin{figure}[b]
	\includegraphics[width=3.4in]{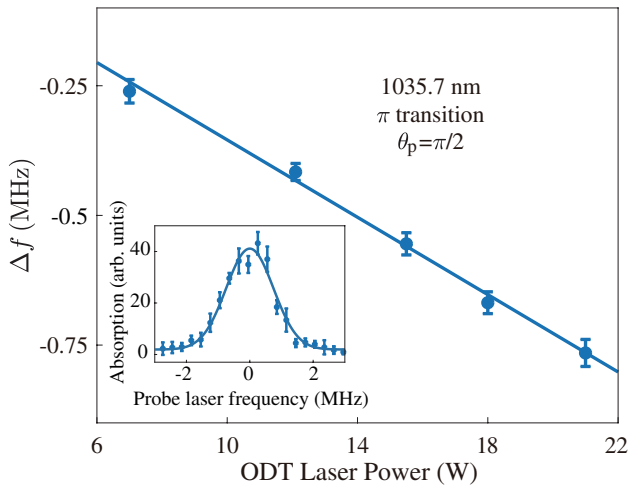}
	\caption {ac Stark shift ($\Delta f$) vs. ODT laser power for the $\pi$ transition at the ODT wavelength of 1035.7 nm. The slope of the fit line is DSSC. The ac Stark shift is determined by the center position of the absorption spectrum (see inset). A Gaussian fit (solid curve) gives a FWHM of 1.8 MHz for an ODT power of 15.5 W. The peak is broadened due to both the ac Stark shift of the ODT and the Zeeman shift in the MOT.}
\label{MW_example10357}
\end{figure}

Absorption-image spectroscopy is used to measure the differential ac Stark shift of the $6s^2{}\,^1S_0-6s6p\,{}^3P_1$ transition. As shown in Fig.~\ref{Apparatus}(b), the 555.8 nm probe laser passing through a dichroic mirror is overlapped with the ODT. Under a uniform $B_{bias}$ of 2 Gauss, the  Zeeman splitting of the ${}^3P_1$ state is 4 MHz and the $\pi$ and $\sigma$ transitions (linewidth 181 kHz) are well separated in the spectrum. For each data point in Fig.~\ref{MW_example10357}, we scan the frequency of 555.8 nm probe laser to record an absorption spectrum and determine the ac Stark shift. For each ODT wavelength, the ac Stark shift has a linear dependence on the ODT power. The slope of this dependence gives the DSSC at this particular ODT wavelength. Fig.~\ref{MW_example10357} shows the DSSC for the $\pi$ transition at 1035.7 nm.

\section{results and discussion} \label{sec:results and discussion}
\begin{figure}[h]
	\includegraphics[width=3.4in]{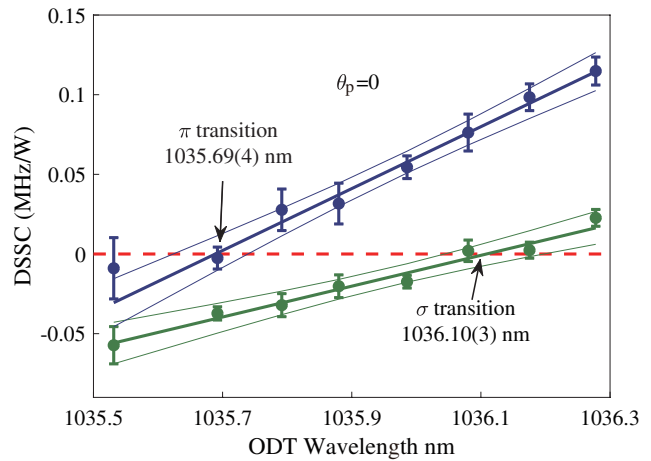}
	\caption {DSSC vs. ODT wavelength. Error bar indicates the standard deviation of each data point and the two thin lines illustrate the confidence bounds (95\%). The zero crossing points are 1036.10(3) nm for the $\sigma$ transition and 1035.69(4) nm for the $\pi$ transition.}
\label{MWresults}
\end{figure}

We measure DSSCs for different ODT wavelengths, which are measured by a calibrated wave meter (Bristol 671A) with an uncertainty of less than 1 ppm (Table~\ref{tab:table2}). All wavelength values are those of wavelengths in vacuum. By linear fitting the data, we determine the zero crossing wavelength for each transition. For the $\sigma$ transition, we average the DSSCs of $\sigma^+$ and $\sigma^-$ transitions to eliminate any residual vector shifts (due to residual circular polarization of the ODT). Under the experimental setting of $\theta_{p}=\pi/2$, the magic wavelengths are 1035.83(3) nm ($\sigma$ transitions, $|\Delta m| = 1$) and 1036.10(3) nm ($\pi$ transition, $\Delta m = 0$). With the corresponding geometric factors, we separate the scalar and tensor parts of DSSCs for $\theta_{p}=\pi/2$ and recombine them for $\theta_{p}=0$. The zero crossing wavelengths for $\theta_{p}=0$, and thus the magic wavelengths are 1036.10(3) nm ($\sigma$ transitions, $|\Delta m| = 1$) and 1035.69(4) nm ($\pi$ transition, $\Delta m = 0$)(Fig.~\ref{MWresults}).

In the experiment, a deviation of $\theta_{p}$ from $\pi/2$ causes a systematic error. We evaluate $\theta_{p}$ using an indirect method. First, the polarizing beam splitter (PBS) that define the ODT polarization is verified to be parallel to another PBS for the 555.8 nm probe laser beam. Second, the angular relation between the polarization of the 555.8 nm probe laser and $B_{bias}$ can be directly measured relying on the fact that the $\pi$ transition is forbidden at $\theta_p = \pi/2$. At the maximum $\sigma$-to-$\pi$ peak ratio of (59$\pm9$), we determine that $\theta_{p}=82.6\pm 0.6^{\circ}$. Other systematic effects are also investigated, and the results are listed in Table~\ref{tab:table2}. The statistical uncertainty dominates in this measurement. The final results are listed in Table~\ref{tab:table1}.

\begin{table}[h]
\caption{\label{tab:table1}%
The measured and calculated values of the magic wavelengths (in vacuum) for the $6s^2\,{}^1S_0-6s6p\,{}^3P_1$ transition in the neutral Yb atom.}
\begin{ruledtabular}
\begin{tabular}{lcccc}
 &\multicolumn{2}{c}{$\theta_{p}=0$}&\multicolumn{2}{c}{$\theta_{p}=\pi/2$}\\
\cline{2-3}  \cline{4-5}
 &$\sigma$&$\pi$&$\sigma$&$\pi$\\
\hline
Experiment\footnotemark[1]&1036.12(3)&1035.68(4)&1035.83(3)&1036.12(3)\\
Theory\footnotemark[1]&1035.9&1035.5&1035.7&1035.9\\
Theory\footnotemark[2]&1036.0(3)&1035.7(2)\\
\end{tabular}
\end{ruledtabular}
\footnotetext[1]{This work.}
\footnotetext[2]{Z.-M Tang et al. (2018) \cite{ZhiYanJun18}.}
\end{table}

\begin{table}[h]
\caption{\label{tab:table2}%
Error evaluation. The wavelength error is given in the unit of “pm” for pico-meter.}
\begin{ruledtabular}
\begin{tabular}{lcccc}
Contribution&\multicolumn{2}{c}{Correction (pm)}&\multicolumn{2}{c}{Uncertainty (pm)}\\
\cline{2-3}  \cline{4-5}
&$\sigma$&$\pi$&$\sigma$&$\pi$\\
\hline
Statistics &0&0&29&38\\
$\theta_p$&+15&-7&2&1\\
ODT power ($5\%$)&0&0&4&4\\
Wave meter &0&0&0.5&0.5\\
Total&+15&-7&29&38
\end{tabular}
\end{ruledtabular}
\end{table}

\section{Outlook} \label{sec: outlook}
By tuning the ODT wavelength to or near the magic wavelength, Sisyphus type narrow-line laser cooling can be investigated to achieve a sub-Doppler temperature. It may also help enhance the MOT-to-ODT transfer efficiency since laser cooling can be applied continuously as the atoms fall into the ODT. Furthermore, since an ODT operating at this magic wavelength provides an environment without ac Stark shift or broadening, for the odd-isotope ${}^{171}$Yb, optical pumping and spin-state detection can be conducted on the narrow $6s^2\,{}^1S_0-6s6p\,{}^3P_1$ transition for spin-polarized cold-atom experiments.

\section*{Acknowledgements} \label{sec:acknowledgements}
We thank M. Bishof, M. R. Dietrich, N. D. Lemke, P. Mueller, and J. T. Singh for helpful discussions. This work has been supported by the National Natural Science Foundation of China (NSFC) through Grants No. 11704368, No. 91636215 and by the Strategic Priority Research Program of the Chinese Academy of Sciences through Grant No. XDB21010200. M. S. Safronova is supported by the National Science Foundation (NSF) of United States through Grant No. PHY-2012068.

\nocite{*}

\end{document}